\documentclass{optica-article}
\journal{opticajournal} 
\articletype{Research Article}
\usepackage{subcaption}
\usepackage{float}
\usepackage{lineno}
\begin{document}
\title{Free-Space Characterization Setup for Low-loss Aluminum Oxide Waveguides at 261 nm}
\author{Vahram Voskerchyan\authormark{1}, Dawson Bonneville\authormark{3}, Lantian Chang\authormark{1} and  S.M. García Blanco\authormark{1,2}}

\address{\authormark{1}Integrated Optical System Group, MESA+ Institute for Nanotechnology, University of Twente, 7500AE Enschede, The Netherlands\\
\authormark{2}Aluvia Photonics, The Netherlands\\
3$^{+}$ Photonics Inc., Canada\\}

\email{v.voskerchyan@utwente.nl}

\begin{abstract*}
We present a methodology for the characterization of deep-ultraviolet (UV) photonic integrated circuits (UV-PICs) based on polycrystalline Al$_2$O$_3$, operating at a wavelength of 261 nm. The platform enables low-loss propagation in the deep UV, and we demonstrate an image-based analysis pipeline for estimating waveguide attenuation using free-space coupling and scattered-light imaging. The characterization approach combines spatial calibration of the imaging system, background analysis, and controlled exposure conditions to extract the exponential decay of scattered light along the propagation direction. Preliminary measurements suggest propagation losses of 4.6 dB/cm for 600 nm wide waveguide, while narrower waveguides exhibit higher attenuation due to increased scattering and reduced mode confinement. This work primarily documents the experimental setup and analysis methodology used for deep-UV characterization, providing a foundation for further validation and refinement of propagation-loss measurements in integrated photonic devices operating in the deep-ultraviolet regime.
\end{abstract*}
\section{Introduction}
Applications operating in the ultraviolet wavelength range (i.e., between 200 and 400 nm), including UV Raman spectroscopy \cite{PC2023}, UV-VIS spectroscopy \cite{Perkampus2013}, microscopy \cite{Lin2022}, metrology \cite{Bennett1988} and quantum computers based on trapped-ions/cold atoms \cite{Mehta2020}, could potentially benefit from the scalability, increased robustness, efficiency, reduced size and cost provided by photonic integration \cite{Blumenthal2020}. Most mature low-loss integrated photonic platforms, such as silicon nitride, cannot operate efficiently in this wavelength range due to their small band gap, leading to prohibitively high waveguide absorption losses below 450 nm \cite{CoratoZanarella2024}.
Alternative material platforms suitable for photonic integrated circuits operating below 400 nm include the III-nitrides (i.e., AlN and AlGaN) \cite{Soltani2016}, silicon oxynitride (SiO$_x$N$_y$) \cite{Mogensen2001,Kervazo2022}, and aluminium oxide (Al$_2$O$_3$) \cite{West2019,He2023,Hendriks2021} thanks to their large bandgap below 200 nm \cite{Soltani2016,Filatova2015}. The high propagation losses demonstrated so far in the AlN platform prevent it yet from serving useful applications in the deep UV wavelength range. On the other hand, the low-refractive index contrast of the SiO$_x$N$_y$ prevents it from reaching the desirable high integration density.
Al$_2$O$_3$ is an excellent material for integrated photonics thanks to its large transparency window from the ultraviolet to mid-IR.
Recently, a 200 mm wafer-scale CMOS-compatible Al$_2$O$_3$ platform using atomic layer deposition (ALD) demonstrated low propagation losses below 0.6 dB/cm in the 360--638 nm range and 4.3--14.7 dB/cm at 266 nm for patterned waveguides \cite{Neutens2025}. In contrast, this work focuses on RF reactive sputtered polycrystalline Al$_2$O$_3$, achieving comparable low losses of $\sim$4.6 dB/cm at 261 nm (for 600 nm width) via free-space image-based characterization, offering a potentially simpler deposition alternative for deep-UV applications.
\section{Channel Waveguide Characterization at 261 nm}
\begin{figure}[H]
    \centering
    \begin{subfigure}{0.7\textwidth}
        \centering
        \includegraphics[width=\textwidth]{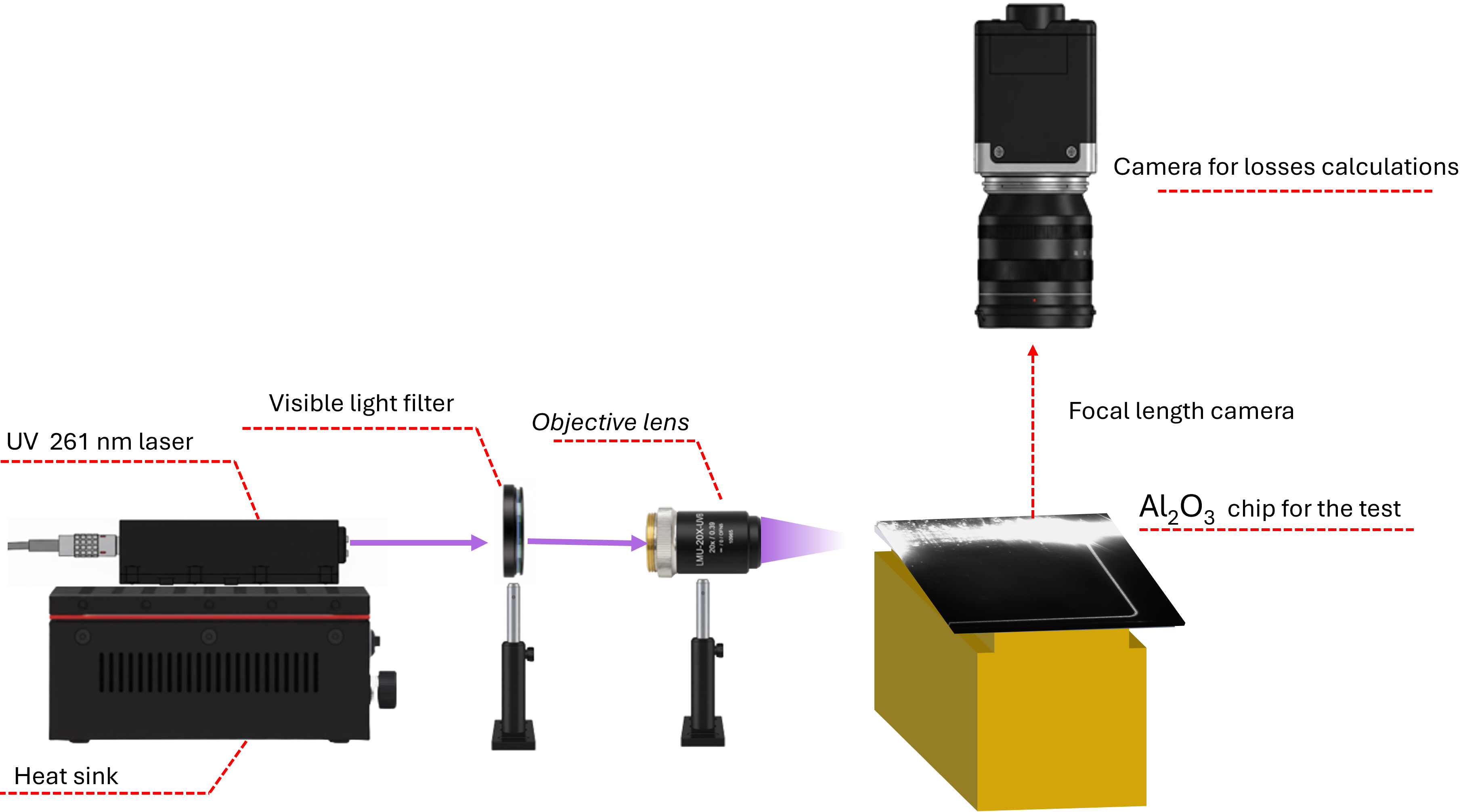}
        \caption{Experimental setup.}
        \label{fig:uv_setup}
    \end{subfigure}
    \vspace{0.5cm}
    \begin{subfigure}{0.4\textwidth}
        \centering
        \includegraphics[width=\textwidth]{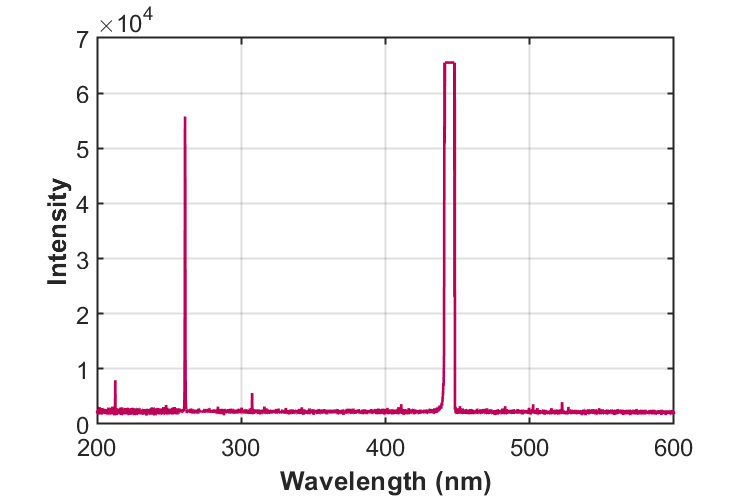}
        \caption{}
        \label{fig:uv_spectrum_before}
    \end{subfigure}
    \begin{subfigure}{0.429\textwidth}
        \centering
        \includegraphics[width=\textwidth]{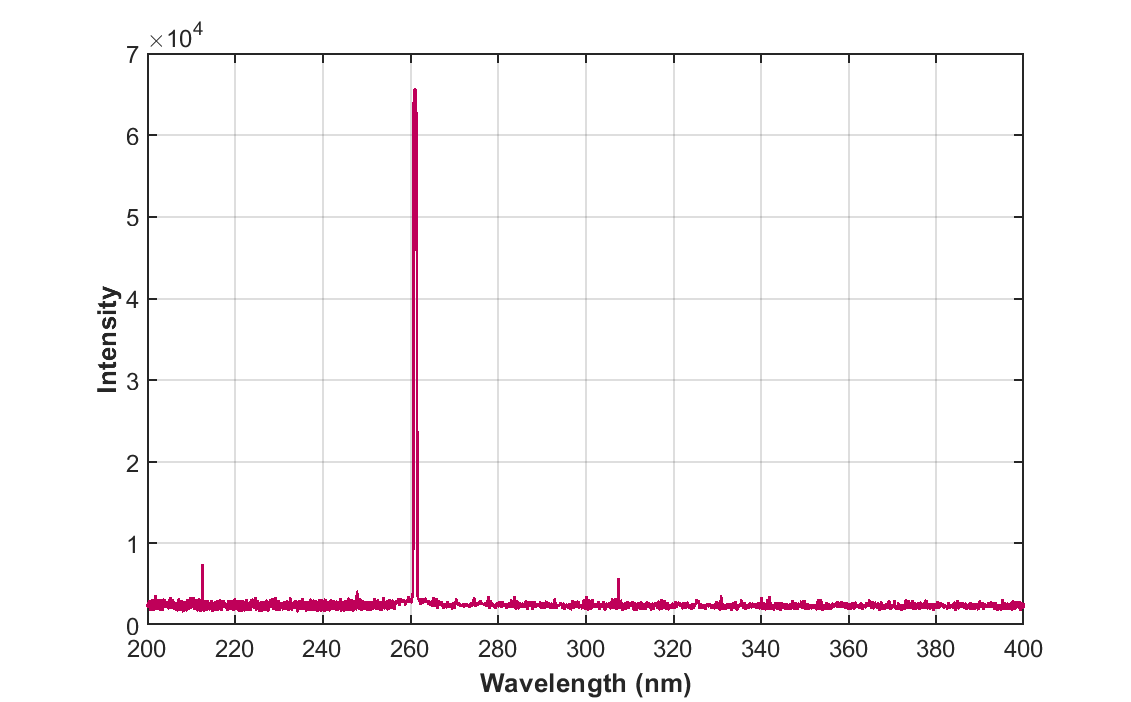}
        \caption{}
        \label{fig:uv_spectrum_after}
    \end{subfigure}
    \caption{Characterization of UV 261 chips: (a) Camera based waveguide loss measurement setup at a wavelength of 261 nm and (b,c) spectrometer data before and after the UV filter.}
    \label{fig:uv_char}
\end{figure}

\begin{figure}[H]
    \centering
    \begin{subfigure}{0.57\textwidth}
        \centering
        \includegraphics[width=\textwidth]{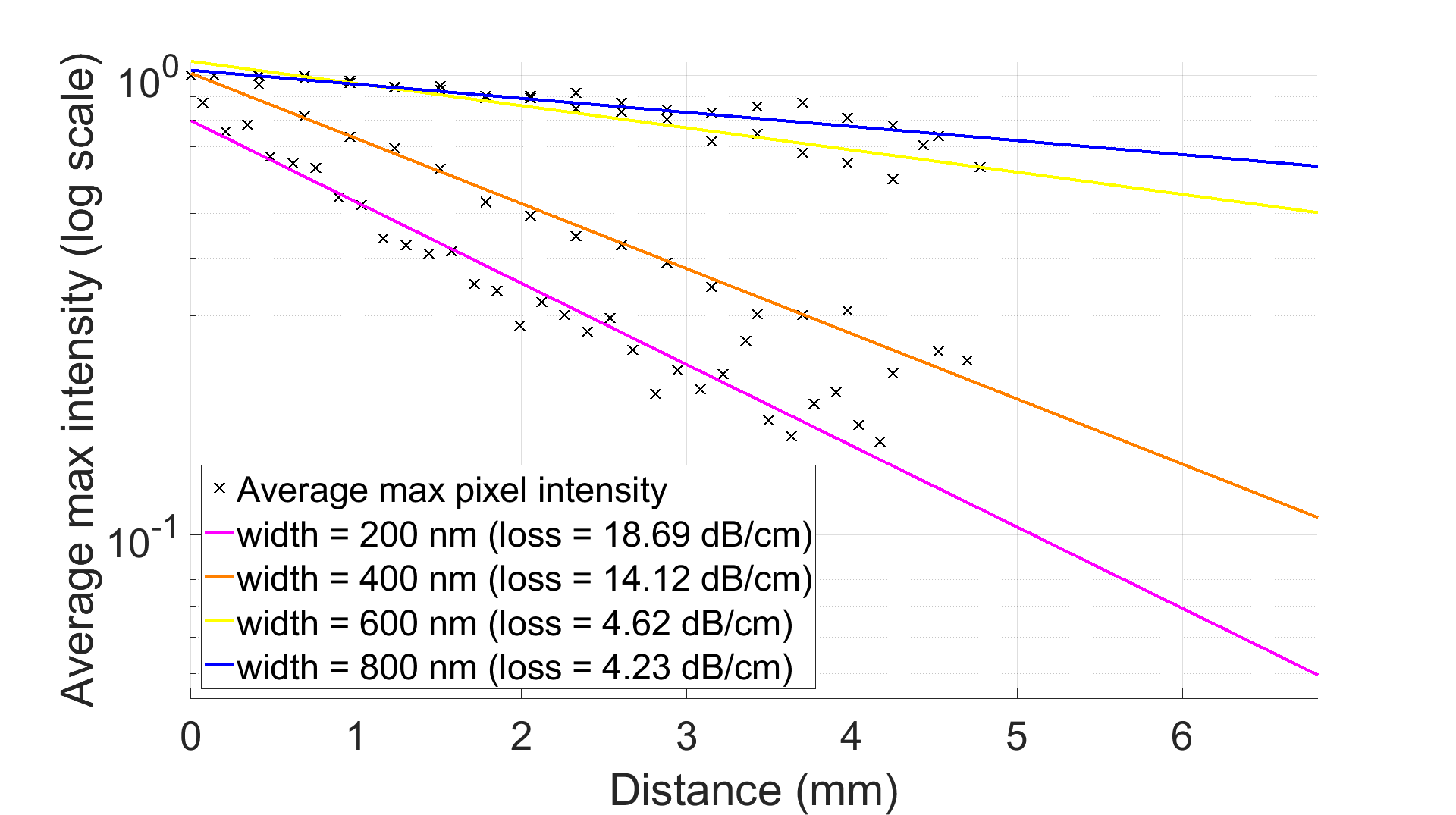}
        \caption{}
        \label{fig:uv261_losses}
    \end{subfigure}
    \hfill
    \begin{subfigure}{0.42\textwidth}
        \centering
        \includegraphics[width=\textwidth]{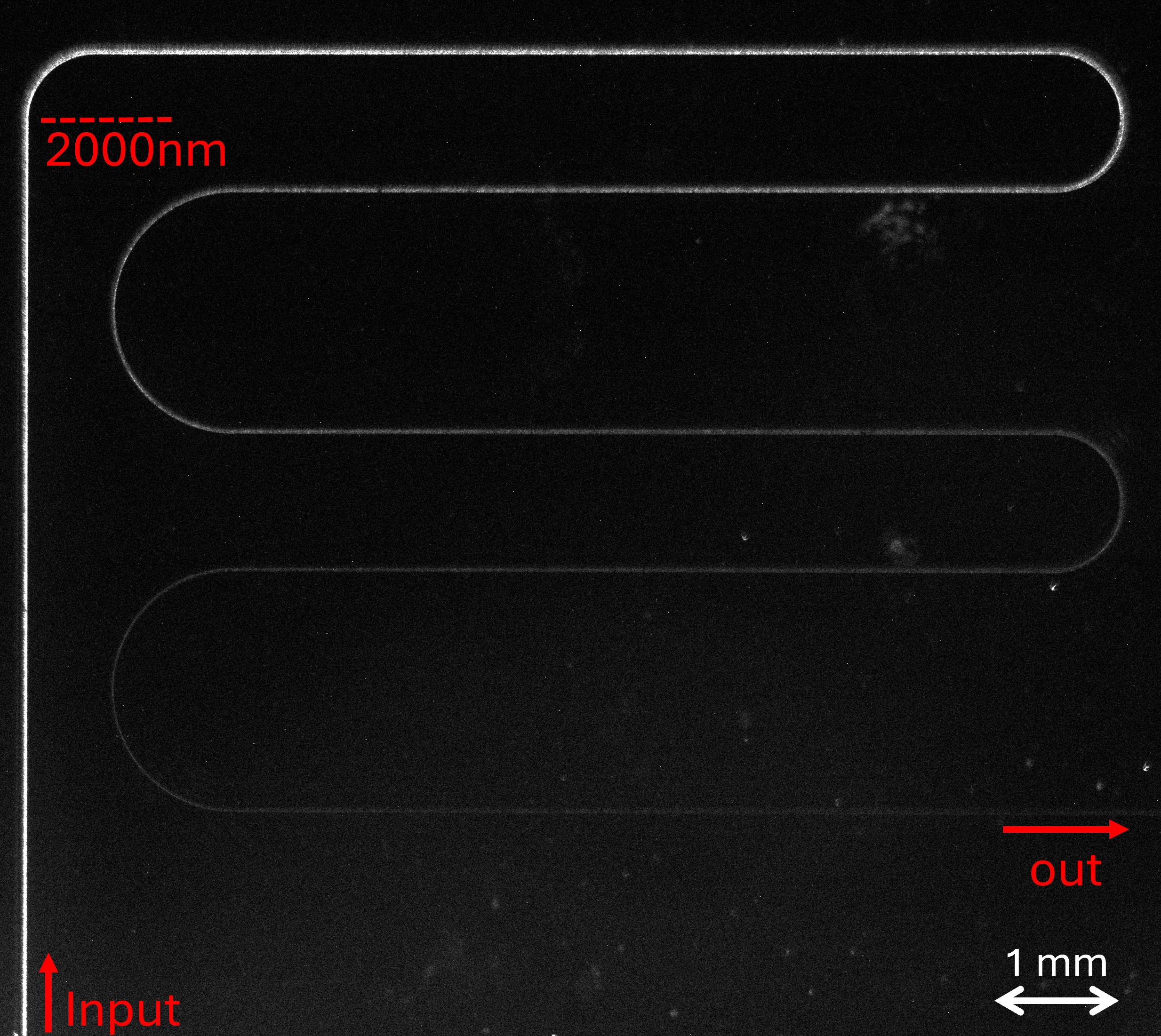}
        \caption{}
        \label{fig:light_prop_2micron}
    \end{subfigure}
    \caption{(a) Measured propagation losses at 261 nm for waveguide widths of 200, 400, 600, and 800 nm. (b) Example of 261 nm light propagation in a 2000 nm wide waveguide.}
    \label{fig:uv261_waveguides}
\end{figure}

Al$_2$O$_3$ layers with a thickness of approximately 95 nm were deposited using an AJA ATC 15000 RF reactive co-sputtering system on 100 mm diameter silicon wafers with an 8 $\mu$m thick thermal oxide layer.
The slab propagation losses were measured using a prism coupler setup (Metricon 2010M) equipped with a fiber loss measurement tool, confirming the low-loss characteristics of the sputtered films in the near-ultraviolet region.
The thickness and refractive index were measured using variable angle spectroscopic ellipsometry (VASE) with a Woollam M-2000UI ellipsometer on a similar Al$_2$O$_3$ layer deposited on a bare silicon substrate.
These optimized Al$_2$O$_3$ films formed the basis for the fabrication of channel waveguides. Electron-beam lithography (EBL) was used to pattern the waveguides. Waveguide structures were etched into the Al$_2$O$_3$ layer using reactive ion etching (RIE). An 8 $\mu$m silicon dioxide (SiO$_2$) cladding was then deposited via low-pressure chemical vapor deposition (LPCVD). To further improve film quality, the wafers were annealed. A top cladding layer of 8 $\mu$m SiO$_2$ was deposited using plasma-enhanced chemical vapor deposition (PECVD). Finally, the photonic integrated circuits (PICs) were diced using a diamond blade (Disco F1230).
The loss characterization of aluminum oxide waveguides at 261~nm was performed using a free-space coupling setup, shown in Fig.~\ref{fig:uv_char}(a). A 261~nm UV laser source was coupled into the waveguides using an objective lens (Thorlabs LMU-20X-UVB), with the chip positioned at the focal point of the objective to achieve optimal coupling efficiency.
Mode-overlap FDE simulations determined the optimal objective lens parameters, yielding an overlap efficiency of approximately 46\%, indicating effective coupling between the objective output and the waveguide core.
The optical chip was positioned on a 5-axis piezoelectric stage for precise alignment. A UV-sensitive monochrome camera was used to record the intensity distribution of the scattered UV light from the waveguides. The captured intensity data were subsequently analyzed to determine propagation losses.
Because UVC light poses significant health hazards, including skin burns and eye damage, the entire experimental setup was fully enclosed within a protective housing. UV-blocking filters were mounted on plexiglass panels to absorb harmful radiation.
To remove residual shorter-wavelength light from the laser, a UV-transmitting filter (UG11-UV) was used, transmitting only wavelengths between 261~nm and 300~nm. At the output facet of the chip, a spectrometer measured the transmitted spectrum, confirming that only 261~nm light propagated during the experiment (see Figure~\ref{fig:uv_char}b).
The waveguides were designed with sufficient spacing to prevent slab light from coupling, ensuring accurate image-based loss calculations.
The enclosed setup incorporated a free-space coupling stage, a piezoelectric stage for precise chip alignment, and a UV-sensitive camera for recording the intensity distribution of the light on the chip. This configuration ensured both operator safety and accurate optical characterization under well-controlled conditions.
The propagation losses were estimated by analyzing the spatial decay of scattered light along the waveguide. The scattered intensity was extracted from the camera images along the propagation direction and fitted with the exponential model $I(x) = I_0 \exp(-\alpha x)$ where $I(x)$ represents the measured intensity at position $x$, $I_0$ is the initial intensity, and $\alpha$ is the propagation loss coefficient.
To improve the robustness of the extraction, the recorded images were segmented into multiple spatial blocks along the waveguide length. For each block, the intensity was averaged across the waveguide region to reduce sensitivity to localized scattering events or pixel-level noise. The resulting averaged intensity values were used to estimate the decay profile along the propagation direction.
The measured intensity values were fitted on a logarithmic scale, and the goodness of fit for the exponential decay model was quantified using the coefficient of determination. For example, the extracted decay for the 200 nm wide waveguide yielded a goodness-of-fit value of $R^2 = 0.95$, indicating a reasonable agreement with the exponential attenuation model expected for waveguide propagation. 
\vspace{-0.6cm}
\begin{figure}[H]
    \centering
    \begin{subfigure}{0.57\textwidth}
        \centering
        \includegraphics[width=\textwidth]{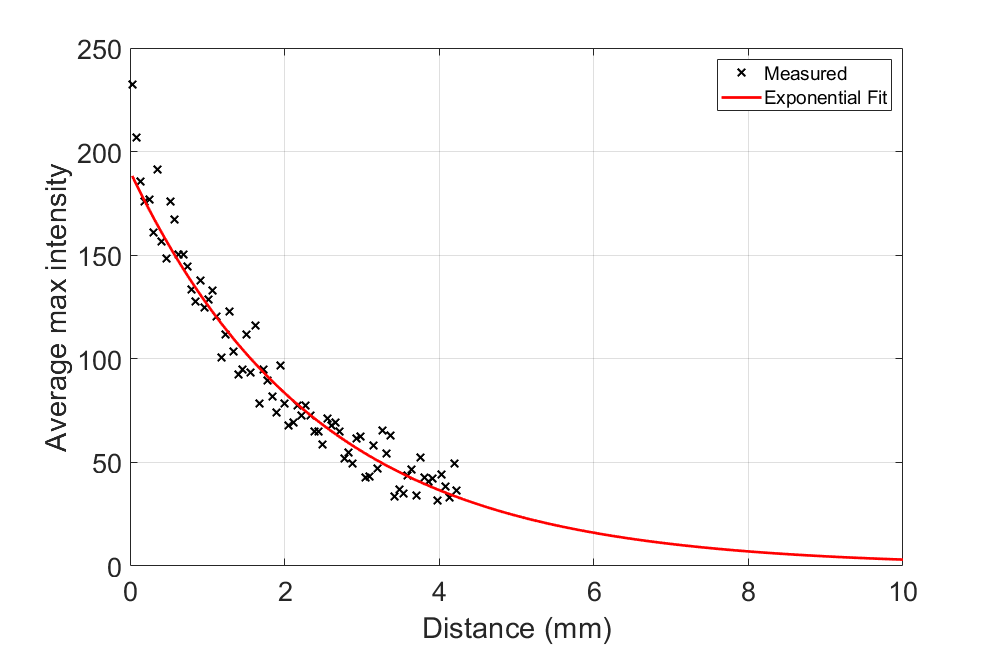}
        \caption{}
        \label{fig:example_loss}
    \end{subfigure}
    \begin{subfigure}{0.1\textwidth}
        \centering
        \includegraphics[width=\textwidth]{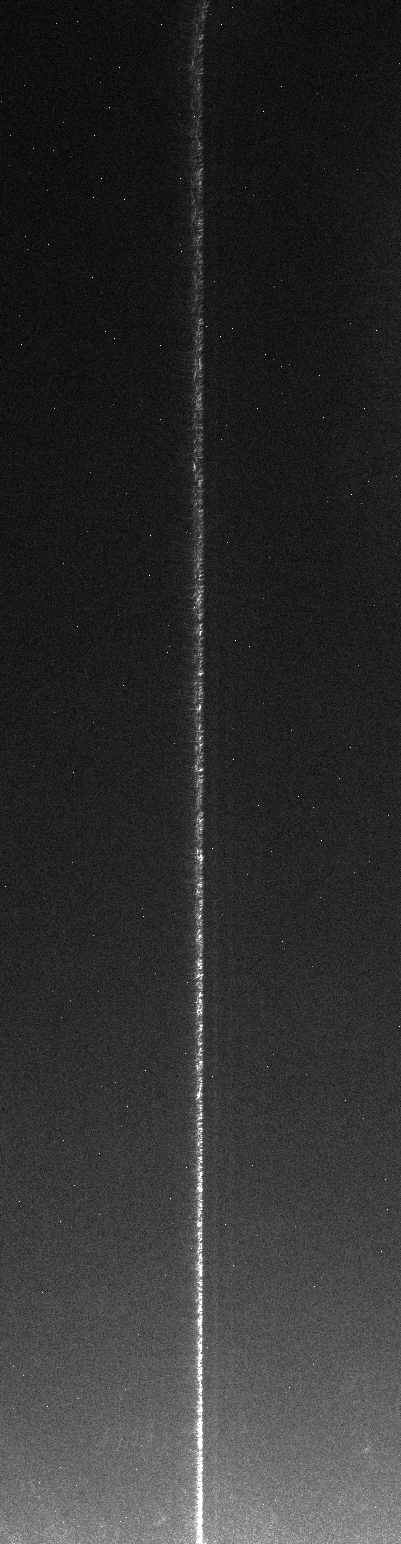}
        \caption{}
        \label{fig:uv200_prop}
    \end{subfigure}
    \caption{(a) Exponential decay of scattered intensity along a 200 nm wide waveguide at 261 nm, with the solid line representing the exponential fit. (b) Captured image showing 261 nm light propagation in the 200 nm waveguide.}
    \label{fig:uv200}
\end{figure}
The corresponding propagation power drop over the waveguide length was determined from the measured intensity as $10 \log_{10} \left(I_\mathrm{max}/I_\mathrm{min}\right)$, ensuring accurate quantification of the attenuation.
The spatial calibration of the propagation length was performed by accounting for the magnification of the imaging system and the physical pixel size of the camera sensor. The objective lens magnification and camera pixel pitch were used to convert distances measured on the sensor plane to the corresponding physical distances on the chip. The analyzed image region was selected to capture the full straight waveguide section, ensuring that the extracted propagation length corresponds to the physical waveguide segment defined in the chip layout.
To avoid distortion of the measured decay profile, camera exposure parameters were selected such that the recorded intensity remained within the linear dynamic range of the UV-sensitive detector. Pixels exhibiting saturation were excluded from the analysis to ensure that the extracted intensity values correspond to the linear response region of the camera. Moreover, the background light also needs to be adjusted, to accurately represent pixel intensity.
The measured losses decrease from approximately 18 dB/cm for the 200 nm wide waveguide to approximately 4.6 dB/cm for the 600 nm wide waveguide, as shown in Fig.~\ref{fig:uv261_waveguides}(a). Measurements were performed for waveguide widths of 200 nm, 400 nm, 600 nm, 800 nm, and 2 $\mu$m.
The observed trend is consistent with increased sidewall scattering and reduced mode confinement in narrower waveguides. As the waveguide width increases, optical confinement improves and the overlap of the optical mode with the etched sidewalls decreases, resulting in lower propagation losses. The design used a minimum bend radius of 450~$\mu$m, including the 200~nm structures.Reduced mode confinement makes bends more sensitive to Rayleigh scattering. Because scattering scales as $1/\lambda^4$, sidewall roughness that is negligible at infrared wavelengths becomes a major loss source. As a result, the attenuation in the 200~nm waveguides arises from both strong straight-waveguide scattering and additional radiative leakage in the 450~$\mu$m bends. 

\begin{figure}[htbp]
    \centering
    \includegraphics[width=0.6\textwidth]{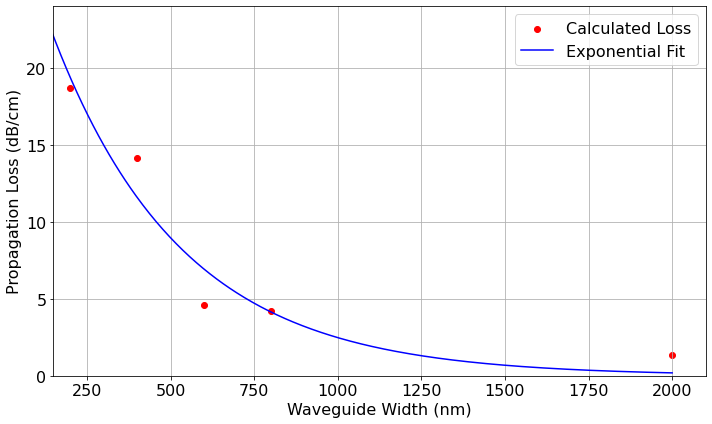}
    \caption{Measured propagation losses at 261~nm across various waveguide widths. The exponential fit is included for visualization purposes only.}
    \label{fig:uv261_all}
\end{figure}
Fig.~\ref{fig:uv261_all} summarizes the measured propagation losses at a UV wavelength of 261 nm for waveguides with different widths. Narrower waveguides (e.g., 200 nm) exhibit significantly higher losses due to increased sidewall scattering and poor mode confinement. As the waveguide width increases, the loss decreases substantially, with the lowest loss observed for the 2000 nm wide waveguide.
\section{Conclusion}
We have presented a free-space characterization methodology for aluminum oxide waveguides operating in the deep-ultraviolet spectral region at 261 nm. The experimental approach combines free-space coupling, UV-sensitive imaging, and image-based analysis of scattered light to estimate propagation losses in integrated waveguides.
Preliminary measurements indicate propagation losses on the order of 4.6 dB/cm for 600 nm wide waveguides, while narrower structures exhibit higher attenuation due to increased scattering. These measurements should be interpreted as an initial estimation obtained from image-based analysis, and further measurements are ongoing to refine the quantitative accuracy of the loss values.
The primary contribution of this work is the documentation of the experimental setup and analysis pipeline for deep-UV photonic characterization. The methodology provides a practical framework for future measurements and further optimization of Al$_2$O$_3$ photonic integrated circuits operating at ultraviolet wavelengths. Independent characterization techniques are currently being investigated to further validate the reported propagation loss values.

\section{Author Contribution}
V.V. built the experimental setup, designed the waveguides, coordinated fabrication with Aluvia Photonics, ordered equipment, integrated the 261 nm UV-C laser, and conceptualized the free-space coupling setup. D.B assisted with initial setup. L.C. offered technical feedback regarding magnification and saturation. S.M.G.B. supervision and funding acquisition.

\end{document}